\documentclass[runningheads]{llncs}

\usepackage{graphicx}
\usepackage{booktabs}
\usepackage{cite}
\usepackage{xcolor}
\usepackage{multirow}
\usepackage{amsfonts}
\usepackage{array}

\usepackage[colorlinks=true, urlcolor=blue, linkcolor=red]{hyperref}

\begin{document}
\title{Xray2Xray: World Model from Chest X-rays with Volumetric Context}
%
%
\author{Zefan Yang\inst{1} \and 
Xinrui Song\inst{1} \and
Xuanang Xu\inst{1} \and
Yongyi Shi\inst{1} \and
Ge Wang\inst{1} \and 
Mannudeep K. Kalra\inst{2} 
\and Pingkun Yan\inst{1}\thanks{Corresponding author: yanp2@rpi.edu}} 
%
\authorrunning{Zefan Yang et al.}
%
\institute{Princeton University, Princeton NJ 08544, USA \and Springer Heidelberg, Tiergartenstr. 17, 69121 Heidelberg, Germany \email{lncs@springer.com}\\ \url{http://www.springer.com/gp/computer-science/lncs} \and ABC Institute, Rupert-Karls-University Heidelberg, Heidelberg, Germany\\ \email{\{abc,lncs\}@uni-heidelberg.de}}
\institute{Department of Biomedical Engineering and Center for Biotechnology and Interdisciplinary Studies, Rensselaer Polytechnic Institute, Troy, NY, USA \and Department of Radiology, Massachusetts General Hospital, Harvard Medical School, Boston, MA, USA}
%
\maketitle              
\begin{abstract}
Chest X-rays (CXRs) are the most widely used medical imaging modality and play a pivotal role in diagnosing diseases. However, as 2D projection images, CXRs are limited by structural superposition, which constrains their effectiveness in precise disease diagnosis and risk prediction. To address the limitations of 2D CXRs, this study introduces Xray2Xray, a novel World Model that learns latent representations encoding 3D structural information from chest X-rays. Xray2Xray captures the latent representations of the chest volume by modeling the transition dynamics of X-ray projections across different angular positions with a vision model and a transition model. We employed the latent representations of Xray2Xray for downstream risk prediction and disease diagnosis tasks. Experimental results showed that Xray2Xray outperformed both supervised methods and self-supervised pretraining methods for cardiovascular disease risk estimation and achieved competitive performance in classifying five pathologies in CXRs. We also assessed the quality of Xray2Xray's latent representations through synthesis tasks and demonstrated that the latent representations can be used to reconstruct volumetric context. 


\keywords{World Model \and Chest X-ray Interpretation \and Computed Tomography.}
\end{abstract}
\section{Introduction}


Chest X-rays (CXRs) are the most commonly used medical imaging modality and play a pivotal role in diagnosing diseases related to the lungs, heart, bones, and blood vessels.
However, as 2D projection images, CXRs are limited by structural superposition along the X-ray paths, which constrains their effectiveness in precise disease diagnosis and risk prediction. 
Although clinical practice commonly orders CXRs from frontal and lateral views to mitigate the issue of superposition \cite{feigin2010lateral}, the amount of 3D spatial information provided by these two views remains highly limited due to the extreme sparsity of projection angles. 
To address the limitations of 2D CXRs, this study introduces Xray2Xray, a novel World Model that enhances CXR analysis by capturing volumetric and contextual representations, achieved by synthesizing new X-ray radiographs aligned with ground-truth computed tomography (CT) volumes.

X-ray projections of a patient's volume from different angular positions inherently encode 3D spatial information, as described by the principles of tomographic reconstruction \cite{suetens2017fundamentals}. Building on this insight, this study proposes training Xray2Xray to model the transition dynamics of X-ray projections, enabling it to capture latent representations of the chest volume for downstream tasks.
The use of multiple X-ray projections to improve prediction accuracy was explored in X-ray dissectography \cite{niu2022multi}, where a network was trained to reconcile multi-view information with 3D spatial features. 
However, due to the limited number of projections, the network failed to capture 3D latent representations.
Additionally, prior studies have leveraged the concept of World Models to learn latent representations of the environment, which were then used to enable agentic decision-making \cite{ha2018world, zhou2024dino, hafner2023mastering, micheli2023transformers}. 
However, these approaches did not incorporate holistic volumetric representations to enhance downstream task predictions.
The proposed Xray2Xray differentiates from related works by introducing a novel transition framework that learns latent representations of volumetric context from X-ray projections and integrating latent features across steps for downstream predictions.

The proposed Xray2Xray consists of a vision model that embeds input projections into latent tokens and a transition model that conducts next-projection prediction. For downstream adaptation, we conditioned Xray2Xray on an initial CXR to generate latent representations that capture 3D spatial information for disease diagnosis.
We trained Xray2Xray using CT volumes from the National Lung Screening Trial (NLST) dataset \cite{national2011reduced} and leveraged it to extract latent representations for cardiovascular disease (CVD) risk prediction on NLST and chest disease diagnosis on CheXpert \cite{irvin2019chexpert}, incorporating domain adaptation to mitigate domain shift. Xray2Xray outperformed both supervised methods (ResNet-50 \cite{he2016deep}, ViT-S \cite{dosovitskiy2021an}, and BI-Mamba \cite{yang2024cardiovascular}) and pretraining methods (CheXFound \cite{yang2025chest} and iGPT \cite{chen2020generative}) that simply combined multiview information without extrapolating 3D context.
On CheXpert, Xray2Xray also demonstrated competitive performance in classifying five pathologies, though it remained behind the discriminative pretraining method CheXFound. 
We further assessed the quality of Xray2Xray's latent representations through projection synthesis tasks and demonstrated that the synthetic X-ray projections were plausible for reconstructing 3D spatial information. 
Overall, Xray2Xray introduces a novel approach for learning 3D latent representations from CXRs, which will enable the design of novel pretraining techniques and downstream applications.






\begin{figure}[t]
\includegraphics[width=\textwidth]{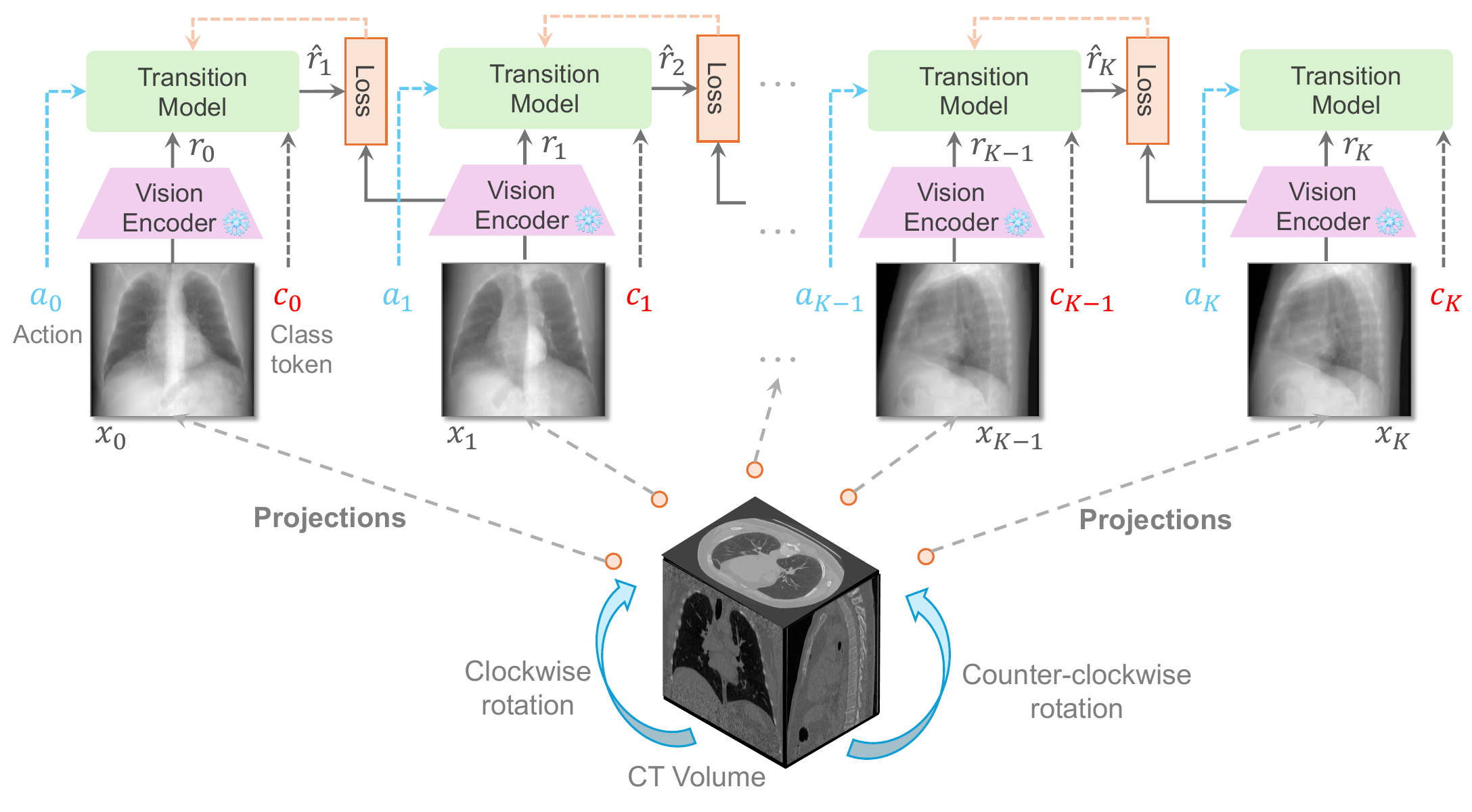}
\caption{Overall framework for Xray2Xray training. Xray2Xray is composed of a vision encoder and a transition model. We trained Xray2Xray with X-ray projections over different angular positions. The vision encoder first converted the input $x_k$ to a latent token sequence $r_k$. The transition model then predicted the next-projection $r_{k+1}$ conditioned on $r_k$. We used a cross-entropy loss to supervise Xray2Xray training.}
\label{fig:framework}
\end{figure}

\section{Method}
The proposed Xray2Xray consists of a vision model that compresses projection images into latent representations and a transition model that captures the dynamics of projection images as the X-ray source moves (Fig. \ref{fig:framework}). This study uses a discrete autoencoder as the vision model, which encodes a projection image into a token sequence and decode it back, and a GPT-style autoregressive Transformer \cite{NIPS2017_3f5ee243} as the transition model, which predicts the next projection image based on the previous step and action. Training Xray2Xray involves two stages. In the first stage, we train a vision model using the VQ-GAN architecture, as detailed in Section \ref{sec:vqgan}. In the second stage, we freeze the VQ-GAN encoder and training a Transformer model to predict the token sequence of the next projection image, as described in Section \ref{sec:tformer}. During inference, Xray2Xray recursively generates next-step predictions based on its own outputs, producing a sequence of class tokens that encode 3D spatial representations for downstream application, as described in Section \ref{sec:infer}.

\subsection{Vision Encoding with Discrete Autoencoder}
\label{sec:vqgan}
The role of the vision model is to encode projection images into a latent feature space, where Xray2Xray learns spatial representations essential for downstream tasks. To achieve this, we use VQ-GAN \cite{esser2021taming} as the vision model to compress projection images into latent features.
We choose VQ-GAN because it captures richer latent features compared to conventional autoencoders, such as variational autoencoders (VAE), thereby preserving more spatial information. Besides, VQ-GAN encodes images into a sequence of tokens, making them well-suited for Transformer models to process and learn 3D spatial representations. 

VQ-GAN tokenizes an image $x$ into a token sequence in two steps. First, the VQ-GAN encoder maps the image into a latent feature map $f \in \mathbb{R}^{N \times C}$. Second, a quantizer converts $f$ into a token sequence $r\in [V]^{N}$. Specifically, the quantizer contains a learnable codebook $Z \in \mathbb{R}^{V \times C}$ with $V$ vectors, where each latent feature $f^{(i, j)}$ is mapped to the index of its nearest code in $Z$.
We train VQ-GAN to reconstruct X-ray projections across a range of angles. The token sequence $r$ generated by our trained VQ-GAN were then used to train the world model to learn 3D representations.


\subsection{Training Xray2Xray for Next-projection Prediction}
\label{sec:tformer}

The training of Xray2Xray involves predicting X-ray projections across a range of angles that satisfy tomographic reconstruction to learn 3D spatial representations. We formulate this process as conditional generative modeling:
\begin{equation}
    p(r_t|a_{t-1}, r_{t-1}, c_{t-1}) = \prod_{i=1}^{N} p(r_{t}^{i}|r_{t}^{<i}, a_{t-1}, r_{t-1}, c_{t-1}).
\end{equation}
The likelihood of the token sequence $r_t = (r_{t}^{1}, ..., r_{t}^{N})$ is conditioned on the token sequence $(a_{t-1}, r_{t-1}, c_{t-1})$. The action token $a_{t-1}$ controls the trainsition direction and the class token $c_{t-1}$ summarizes the conditional information. We leverage the autoregressive structure of Transformer for generation by simply prepending $(a_{t-1}, r_{t-1}, c_{t-1})$ to $r_{t}$. Due the autoregressive process, the likelihood of the $i$-th token $r_{t}^{i}$ is also conditioned on $r_{t}^{<i}$.

Xray2Xray is trained on the token sequences $(r_{0}, ..., r_{K})$ which represent tokenized X-ray projections over $K+1$ angles produced by the VQ-GAN encoder. 
The training objective is to minimize the negative log-likelihood of the next-step token sequence:
\begin{equation}
\mathcal{L}_{Xray2Xray} = - \frac{1}{K} \sum_{t=1}^{K} \log p(r_t|a_{t-1}, r_{t-1}, c_{t-1}).
\end{equation}
This training approach is similar to generative pretraining \cite{yu2021vector, chen2020generative}, with the key distinction that we employ conditional image modeling over a long trajectory to encode latent features representing 3D spatial information.

\subsection{Extracting Latent Representations for Downstream Tasks}
\label{sec:infer}
At inference time, the pretrained Xray2Xray is used for CXR analysis.
To generate latent projection predictions over a trajectory that contain 3D information, Xray2Xray first predicts the subsequent sequence $\hat{r}_1$ given an initial input $r_0$ and then recursively generates the remaining sequences $(\hat{r}_2,...,\hat{r}_K)$ according to $p(\hat{r}_{t}|a_{t-1}, \hat{r}_{t-1},c_{t-1})$. From these sequences $(r_0, \hat{r}_1,....,\hat{r}_K)$, class tokens $(c_0, ..., c_K)$ are extracted for downstream adaptation:
\begin{equation}
y = g(c_0, ..., c_K).
\end{equation}
We train a classifier $g(\cdot)$ on top of the frozen class token sequence to predict the target $y$. The classifier uses an attentional module to integrate latent features, following the design of GLoRI \cite{yang2025chest}. Because Xray2Xray is trained with image modeling objectives, its final-layer features are useful for image generation but may not be optimal for classification. In accordance with related works in generative pretraining \cite{chen2020generative, yu2021vector}, we used the middle-layer features for the classification task. We also verified the quality of the learned representation by reconstructing a CT volume from a frontal or lateral CXR. 


\section{Experiments and Results}

\subsection{Datasets}
We trained Xray2Xray using 1,000 CT volumes from the National Lung Screening Trial (NLST) dataset. To generate X-ray projections, we first positioned the X-ray source at $0^\circ$ to obtain an anterior-posterior frontal projection, and then rotated it counter-clockwise to $-90^\circ$ and clockwise to $90^\circ$ to collect the remaining views. With a parallel beam, the projections spanning $[-90, 90]$ satisfy the requirements for tomographic reconstruction. The X-ray source was rotated in equidistant steps of $5^\circ$, resulting in 37 X-ray projections per CT volume. We partitioned the dataset at patient level into training, validation, and test sets with a 70:10:20 ratio. 

For downstream adaptation, we applied Xray2Xray to two tasks: cardiovascular disease (CVD) risk estimation on the NLST dataset and disease diagnosis on the CheXpert dataset. For CVD risk estimation, we used labels annotated by Chao et al. \cite{chao2021deep} with 7,268 subjects for training, 1,042 subjects for validation, and 2,085 subjects for test. 
For disease diagnosis, we used labels provided by CheXpert \cite{irvin2019chexpert}, containing 191,027, 202, and 518 CXRs for training, validation, and test, respectively.


\subsection{Experimental Details}
We trained VQ-GAN to tokenize X-ray projections from $-90^\circ$ to $90^\circ$, using a codebook size of 1024, a code dimension of 256, and an input resolution of 224. We trained VQ-GAN for 100 epochs using the Adam optimizer with a learning rate of 4.5e-6 and a batch size of 12. We trained an autoregressive Transformer on top of the VQ-GAN encoder with a latent output of $14\times14$ tokens. We set the default embedding dimension to $D=1024$ for Transformer. We used a $2\times D$ embedding table for actions that control clockwise and counter-clockwise rotation and a $K\times D$ embedding table for class tokens corresponding to the $K$ inputs over the trajectory. We trained the Transformer for 100 epochs using the AdamW optimizer with a learning rate of 4.5e-6 and a batch size of 12. We trained classifiers for 10 epochs using the AdamW optimizer. We did hyperparameter search on the validation set for the learning rate in $\{$1e-5, 2e-5, 5e-5, 1e-4, 2e-4, 5e-4, 1e-3, 2e-3, 5e-3$\}$.

We used the area under the receiver operating characteristic curve (AUROC) to evaluate performance on classification tasks. We also evaluated the representation quality of Xray2Xray on image reconstruction and synthesis tasks using the peak signal-to-noise ratio (PSNR) and structural similarity index measure (SSIM) metircs. PSNR measures the loss of signal and SSIM can assess structural similarity.

\begin{table}[t]
    \caption{Downstream adaptation of Xray2Xray for CVD risk estimation in AUROC. $p$-value is computed with DeLong’s algorithm \cite{sun2014fast} to compare the statistical difference between AUROC values. The value in boldface indicates the best result.}
    \label{tab:cvd}
    \centering
\scalebox{0.97}{
    \begin{tabular}{c|c|c|c|c|c}
    \toprule
    Method & Model & $\#$ Param. & Two-view Features Extraction & AUROC & $p$-value \\ \midrule
    \multirow{3}{*}{Supervised} & ResNet-50 \cite{he2016deep} & 25M & Two-view output concat. & 0.775 & 2e-5 \\
    & ViT-S \cite{dosovitskiy2021an} & 21M & [CLS] tokens concat. & 0.748 & 4e-6 \\
    & BI-Mamba \cite{yang2024cardiovascular} & 25M & Two-view input concat. & 0.795 & 0.1326 \\ \midrule
    \multirow{3}{*}{Pretraining} & CheXFound \cite{yang2025chest} & 307M & [CLS] tokens concat.  & 0.757 & 4e-6 \\
     & iGPT \cite{chen2020generative} & 455M & Two-view output concat.  & 0.736 & 5e-8 \\
    & Xray2Xray & 391M & 3D latent representations & \textbf{0.809} & - \\
    \bottomrule
    \end{tabular}}
\end{table}

\begin{table}[t]
    \caption{Downstream adaptation of Xray2Xray for disease diagnosis on the CheXpert dataset. The value in boldface indicates the best result.}
    \label{tab:chexpert}
    \centering
\scalebox{0.96}{
    \begin{tabular}{c|c|c|c|c|c|c}
    \toprule
    & \footnotesize{Atelectasis} & \footnotesize{Cardiomegaly} & \footnotesize{Consolidation} & \footnotesize{Edema} & \footnotesize{Pleural Effusion} & \footnotesize{Mean} \\ \midrule
     ConvNeXt \cite{liu2022convnet} & 0.722 & 0.834 & 0.850 & 0.865 & 0.901 & 0.834 \\ \midrule
     CheXFound \cite{yang2025chest} & 0.748 & 0.882 & 0.905 & 0.891 & 0.931 & \textbf{0.871} \\
     iGPT \cite{chen2020generative} & 0.692 & 0.791 & 0.821 & 0.805 & 0.841 & 0.790 \\
     Xray2Xray  & 0.741 & 0.842 & 0.864 & 0.874 & 0.918 & 0.843 \\
    \bottomrule
    \end{tabular}}
\end{table}

 
\begin{table}[t]
    \caption{Overall reconstruction and synthesis performance. Results are the average metrics computed across the projections at different angular positions.}
    \label{tab:rec}
    \centering
    \begin{tabular}{c|c|c|c|c|c}
    \toprule
     Model architecture & Task & $\#$ Param. & Conditional input & PSNR (dB) & SSIM \\ \midrule
     VQ-GAN \cite{yu2021vector} & Recon. & 67M & - & 30.24$\pm$2.25 & 0.864$\pm$0.047 \\ \midrule
     \multirow{2}{*}{Xray2Xray-depth12} & \multirow{2}{*}{Synthesis} & \multirow{2}{*}{154M} & Lateral view & 15.46$\pm$5.12 & 0.623$\pm$0.162 \\
     & & & Frontal view & 16.27$\pm$3.58 & 0.644$\pm$0.110 \\ 
     \midrule
    \multirow{2}{*}{Xray2Xray-depth24} & \multirow{2}{*}{Synthesis} & \multirow{2}{*}{391M} & Lateral view & 20.63$\pm$2.84 & 0.741$\pm$0.055 \\
     & & & Frontal view & 22.03$\pm$3.66 & 0.762$\pm$0.071 \\
    \bottomrule
    \end{tabular}
\end{table}

\subsection{Results of Xray2Xray on Downstream Applications}
We evaluated the effectiveness of Xray2Xray's representations for downstream applications on the NLST and CheXpert datasets. On NLST, Xray2Xray extracted 3D spatial representations from frontal and lateral radiographs for CVD risk estimation. To be specific, given radiographs from two views, Xray2Xray separately generated the class tokens of projections along the full trajectory. We then concatenated the class tokens from two independent trajectories at each angular position and trained a classifier for CVD risk estimation. 
We compared Xray2Xray with supervised training methods (ResNet-50 \cite{he2016deep}, ViT-S \cite{dosovitskiy2021an}, and BI-Mamba \cite{yang2024cardiovascular}) and self-supervised pretraining methods (CheXFound \cite{oquab2023dinov2} and iGPT \cite{chen2020generative}), as shown in Table \ref{tab:cvd}.
Xray2Xray outperformed ResNet-50, ViT-S, and BI-Mamba by 3.4\% ($p<$0.001), 6.1\% ($p<$0.001), and 1.4\% ($p$=0.1326). Xray2Xray also outperformed pretrained models CheXFound and iGPT by 5.2\% ($p<$0.001) and 7.3\% ($p<$0.001). Xray2Xray's improved performance is attributed to its capability to generate 3D latent representations given frontal and lateral radiographs in contrast to conventional methods that only encoded representations of 2D inputs without extrapolating 3D context.

For disease diagnosis on CheXpert, since there exists a domain gap between the real-world and simulated CXRs, we first used CycleGAN \cite{zhu2017unpaired} to transfer the domain of input CXRs. We then used Xray2Xray to extracted their 3D latent representations. 
As shown in Table \ref{tab:chexpert}, Xray2Xray outperformed the supervised training method ConvNeXt \cite{liu2022convnet} by 0.9\% and the generative pretraining method iGPT \cite{chen2020generative} by 5.3\%. Xray2Xray achieved lower but competitive performance compared with CheXFound \cite{yang2025chest}, exhibiting a performance gap of 2.8\%. CheXFound pretrained a foundation model using DINOv2 \cite{oquab2023dinov2} to learn discriminative class embeddings, which was not enforced in Xray2Xray. Besides, domain shift is another potential issue that hampers Xray2Xray's performance. In future work, we will explore novel approaches to improve Xray2Xray's generalizability.

\begin{figure}[t]
\includegraphics[width=\textwidth]{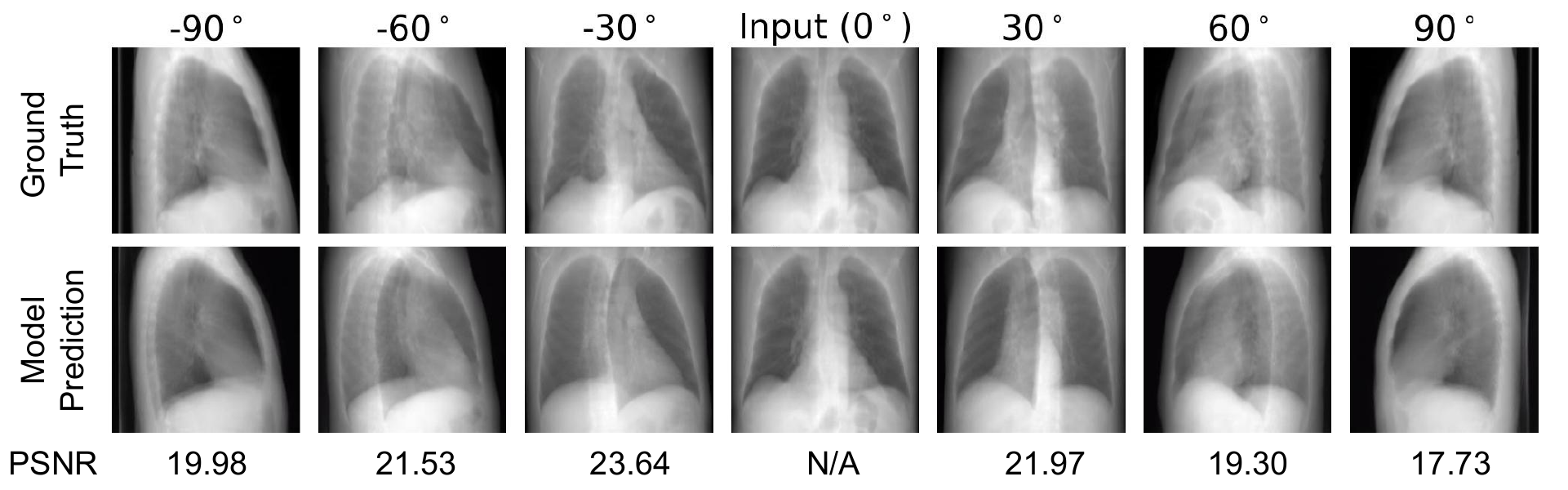}
\caption{Xray2Xray synthesis results. Given an frontal-view projection ($0^\circ$) and actions as inputs, Xray2Xray synthesized the remaining projections across different angular positions without access to extra information. We used a step size of 30 degrees for illustration.}
\label{fig:syn}
\end{figure}

\begin{figure}[t]
\centering
\includegraphics[width=0.85\textwidth]{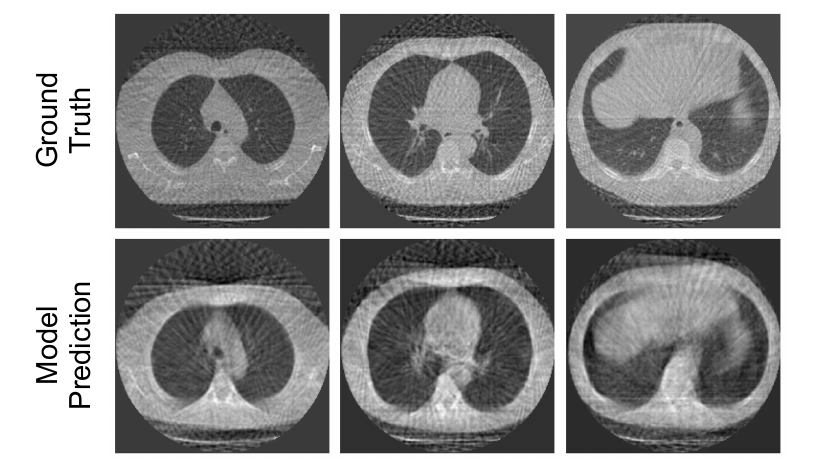}
\caption{Tomographic reconstruction with Xray2Xray's synthetic projections. Top: Ground-truth reconstruction results at different positions along the $z$ axis. Bottom: Reconstruction results with synthetic projections. We used the filtered back projection algorithm to reconstruct tomographic images from projections with a step size of 5 degrees. }
\label{fig:recon}
\label{fig:drrs}
\end{figure}

\subsection{Xray2Xray's Representations for Reconstructing Volumetric Context}
Besides downstream diagnosis tasks, we also evaluated the representation quality of Xray2Xray through reconstruction and synthesis tasks.
We first assessed the VQ-GAN's ability to compress an input into a latent token sequence by examining its reconstructed images.
Table \ref{tab:rec} showed that VQ-GAN's reconstructed images exhibited low noise levels with a PSNR of 30.24 dB and preserved structural information with a SSIM of 0.864.
We then evaluated whether Xray2Xray learned latent features that effectively represent X-ray projections across different angular positions. We conditioned Xray2Xray on an initial input and recursively synthesized token sequences of X-ray projections over the range of [-90, 90] degrees. These token sequences were then decoded by the VQ-GAN decoder into the image space (Fig. \ref{fig:syn}). As shown in Table \ref{tab:rec}, Xray2Xray with a 24-layer Transformer obtained improved latent representations for projection synthesis than the 12-layer version. Xray2Xray's synthetic X-ray projections can be used for tomographic reconstruction (Fig. \ref{fig:recon}), demonstrating Xray2Xray's capability to capture 3D latent representations.








\section{Conclusion}
In summary, this study introduces Xray2Xray, a novel world model that learns latent representations from X-ray projections to capture 3D spatial information.
On downstream tasks, we trained a classifier on top of the 3D latent representations extracted by Xray2Xray given input CXRs.
Experimental results showed that Xray2Xray outperformed both supervised methods and self-supervised pretraining methods for CVD risk estimation and achieved competitive performance in classifying five pathologies in CXRs.
We also demonstrated that the latent representations of Xray2Xray can be used to reconstruct volumetric context, although information loss relative to the ground-truth reconstruction results was observed.
In future work, we will explore novel pretraining approaches to enhance the generalizability as well as the capability to encode 3D latent representations of Xray2Xray.

%
%
%
\bibliographystyle{splncs04}
\bibliography{refs}

\end{document}